\documentclass[prx, preprintnumbers, amsmath, amssymb, superscriptaddress, twocolumn]{revtex4-1}
\usepackage{graphicx, color, dcolumn, natbib, bm, amssymb, amsmath,color}
\graphicspath{{fig/}}
\usepackage[final=true]{hyperref}
\usepackage[normalem]{ulem}

\begin{document}
\title{Gradual pressure-induced change in the magnetic structure of the non-collinear antiferromagnet Mn$_3$Ge}
\author{A. S. Sukhanov}
\affiliation{Max Planck Institute for Chemical Physics of Solids, D-01187 Dresden, Germany}
\affiliation{Institut f{\"u}r Festk{\"o}rper- und Materialphysik, Technische Universit{\"a}t Dresden, D-01069 Dresden, Germany}
\author{Sanjay Singh}
\affiliation{Max Planck Institute for Chemical Physics of Solids, D-01187 Dresden, Germany}
\affiliation{School of Materials Science and Technology, Indian Institute of
Technology (Banaras Hindu University), Varanasi-221005, India}
\author{L. Caron}
\affiliation{Max Planck Institute for Chemical Physics of Solids, D-01187 Dresden, Germany}
\author{Th. Hansen}
\affiliation{Institut Laue-Langevin, 71 avenue des Martyrs, Grenoble Cedex 9, F-38042, France}
\author{A. Hoser}
\affiliation{Helmholtz-Zentrum Berlin f{\"u}r Materialien und Energie, D-14109 Berlin, Germany}
\author{V.\phantom{ }Kumar}
\affiliation{Max Planck Institute for Chemical Physics of Solids, D-01187 Dresden, Germany}
\author{H. Borrmann}
\affiliation{Max Planck Institute for Chemical Physics of Solids, D-01187 Dresden, Germany}
\author{A. Fitch}
\affiliation{European Synchrotron Radiation Facility, 6 rue Jules Horowitz, F-38000 Grenoble, France}
\author{P. Devi}
\affiliation{Max Planck Institute for Chemical Physics of Solids, D-01187 Dresden, Germany}
\author{K. Manna}
\affiliation{Max Planck Institute for Chemical Physics of Solids, D-01187 Dresden, Germany}
\author{C. Felser}
\affiliation{Max Planck Institute for Chemical Physics of Solids, D-01187 Dresden, Germany}
\author{D. S. Inosov}
\affiliation{Institut f{\"u}r Festk{\"o}rper- und Materialphysik, Technische Universit{\"a}t Dresden, D-01069 Dresden, Germany}
\date{\today}
\begin{abstract}

By means of powder neutron diffraction we investigate changes in the magnetic structure of the coplanar non-collinear antiferromagnet Mn$_3$Ge caused by an application of hydrostatic pressure up to 5\phantom{ }GPa. At ambient conditions the kagom\'e layers of Mn atoms in Mn$_3$Ge order in a triangular 120$^{\circ}$ spin structure. Under high pressure the spins acquire a uniform out-of-plane canting, gradually transforming the magnetic texture to a non-coplanar configuration. With increasing pressure the canted structure fully transforms into the collinear ferromagnetic one. We observed that magnetic order is accompanied by a noticeable magnetoelastic effect, namely, spontaneous magnetostriction. The latter induces an in-plane magnetostrain of the hexagonal unit cell at ambient pressure and flips to an out-of-plane strain at high pressures in accordance with the change of the magnetic structure.

\end{abstract}

\maketitle

\section{Introduction}\label{sec:I}
Despite being known for a long time, magnetic materials with non-collinear and non-coplanar spin structures recently attracted giant attention due to the discovery of novel phenomena that can be understood from the point of view of topology. The key concepts, which unify many phenomena in condensed matter previously thought to be unrelated, are the Berry phase and the Berry curvature \cite{Berry}. Among these are the electric polarization, the orbital magnetisation, the anomalous thermoelectric effect, magnetotransport properties, and others (\cite{Rev1} and refs. therein). Separately, one can mention the anomalous Hall effect (AHE). This is a contribution to the transverse conductivity that does not scale with the applied magnetic field. The AHE was first observed in ferromagnets (FMs) more than a hundred years ago, where it was claimed to be an order of magnitude larger than the ordinary Hall effect in non-magnetic conductors. It was later noticed that the transverse conductivity in FMs is proportional to the net magnetisation and remains constant once the saturation is reached. For that reason the AHE in antiferromagnets (AFMs)  was deemed to be forbidden because of compensation of the magnetic moments (\cite{Rev2} and refs. therein).

The non-collinear AFMs Mn$_3$Ge and Mn$_3$Ir were the first AFMs predicted to show the AHE \cite{T1,T2,T5}. The predictions were based on calculations of the Berry phase curvature which was found to be non-vanishing in the presence of a 120$^{\circ}$ triangular magnetic structure.  Shortly afterwards, the large AHE was experimentally found in Mn$_3$Ge \cite{E1,E2} and Mn$_3$Sn \cite{E3}. More generaly, \textit{ab initio} calculations showed that the AHE and the spin Hall effect (SHE) may be present in a series of compounds that share the same magnetic structure: Mn$_3X$ ($X$ = Ge, Sn, Ga, Ir, Rh and Pt) \cite{T3}.

Significance of the AHE and the SHE in AFMs is caused by their potential application in antiferromagnetic spintronics---a new active research area---where magnetic materials free of stray fields are used to create devices capable to manipulate the spin currents \cite{Rev3,T6}. A particular realisation of these ideas was recently proposed for example in Mn$_3$Ge \cite{T4}.

The compound Mn$_3$Ge has a hexagonal crystal structure (space group $P6_3/mmc$, No.\phantom{ }194). The structure consists of 6 Mn atoms and 2 Ge atoms, which occupy the $6h$ and the $2c$ Wyckoff positions, respectively. Magnetic atoms form a kagom\'e lattice with Ge atoms placed at the centers of its hexagonal voids. The kagom\'e layers are densely stacked along the $c$-axis, and 2 layers form the unit cell. As a result, the distance between two neighbouring Mn atoms in the $ab$ plane is very close to the distance between Mn atoms in adjacent layers. Below the N\'eel temperature, $T_{\text{N}} \approx$ 380 K \cite{E1,E2}, Mn$_3$Ge orders in the triangular 120$^{\circ}$ AFM structure that is described by the magnetic space group $Pcm'm'$ \cite{N1,N2}. In this structure, the spins are pointing along $\langle110\rangle$ directions in an inverse manner: $\phi_{i+1} = \phi_{i} - 120^{\circ}$, where $\phi$ is a plain angle. Inversion-related Mn atoms from adjacent layers have their magnetic moments oriented in parallel. The same magnetic structure is found in the closely-related compounds Mn$_3$Sn \cite{N3,N4} and Mn$_3$Ga \cite{N5}.

It is of particular importance to learn how this type of magnetic structure changes under varying conditions, such as magnetic and non-magnetic atomic substitutions, geometric constrains (thin films or nano-structuring), hydrostatic pressure or uniaxial stress.

In this manuscript we study the magnetic structure of Mn$_3$Ge under hydrostatic pressure by means of powder neutron diffraction. The manuscript is organized as follows: in Section \ref{sec:II} we describe the details of the conducted experiments and show the typical data collected. In Section \ref{sec:III} we discuss the results of the data analysis and demonstrate the gradual change in the magnetic structure. Section \ref{sec:IV} is dedicated to correlations seen between the change in the magnetic structure under hydrostatic pressure and a sizeable magnetoelastic effect found in the compound. Finally, in Section \ref{sec:V} we summarize the results.

\section{Powder neutron diffraction measurements}\label{sec:II}

\begin{figure}[t]
        \begin{minipage}{0.99\linewidth}
        \center{\includegraphics[width=1\linewidth]{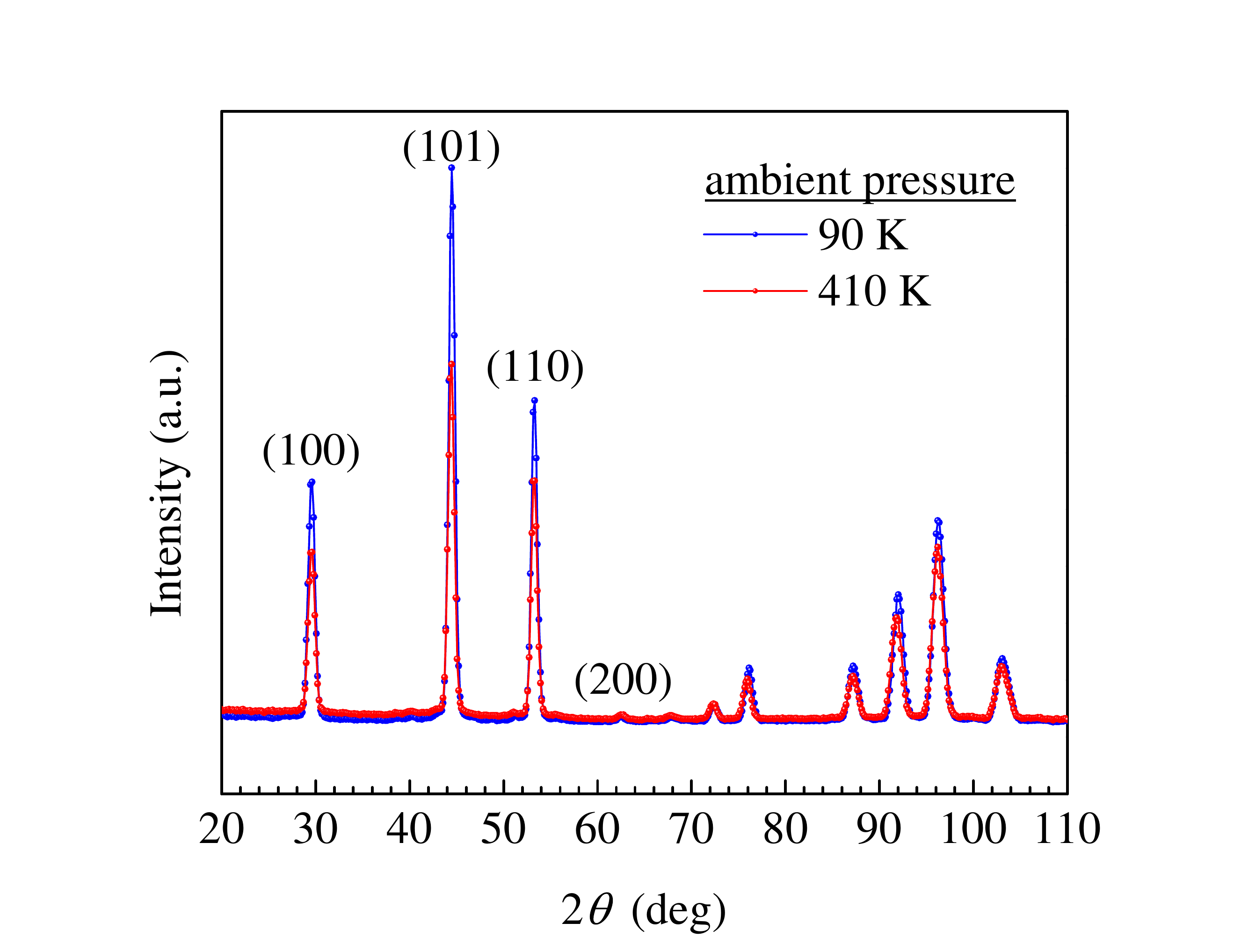}}
        \end{minipage}
        \caption{(color online). Typical diffraction patterns collected at the E6 diffractometer at ambient pressure. The pattern at 410 K corresponds to the paramagnetic state and shows only nuclear contribution, whereas the low-temperature diffractogram contains intensity from both nuclear and magnetic structures. the first four Bragg peaks are denoted by the corresponding indices.}
        \label{ris:fig1}
\end{figure}

\begin{figure}[t]
        \begin{minipage}{0.99\linewidth}
        \center{\includegraphics[width=1\linewidth]{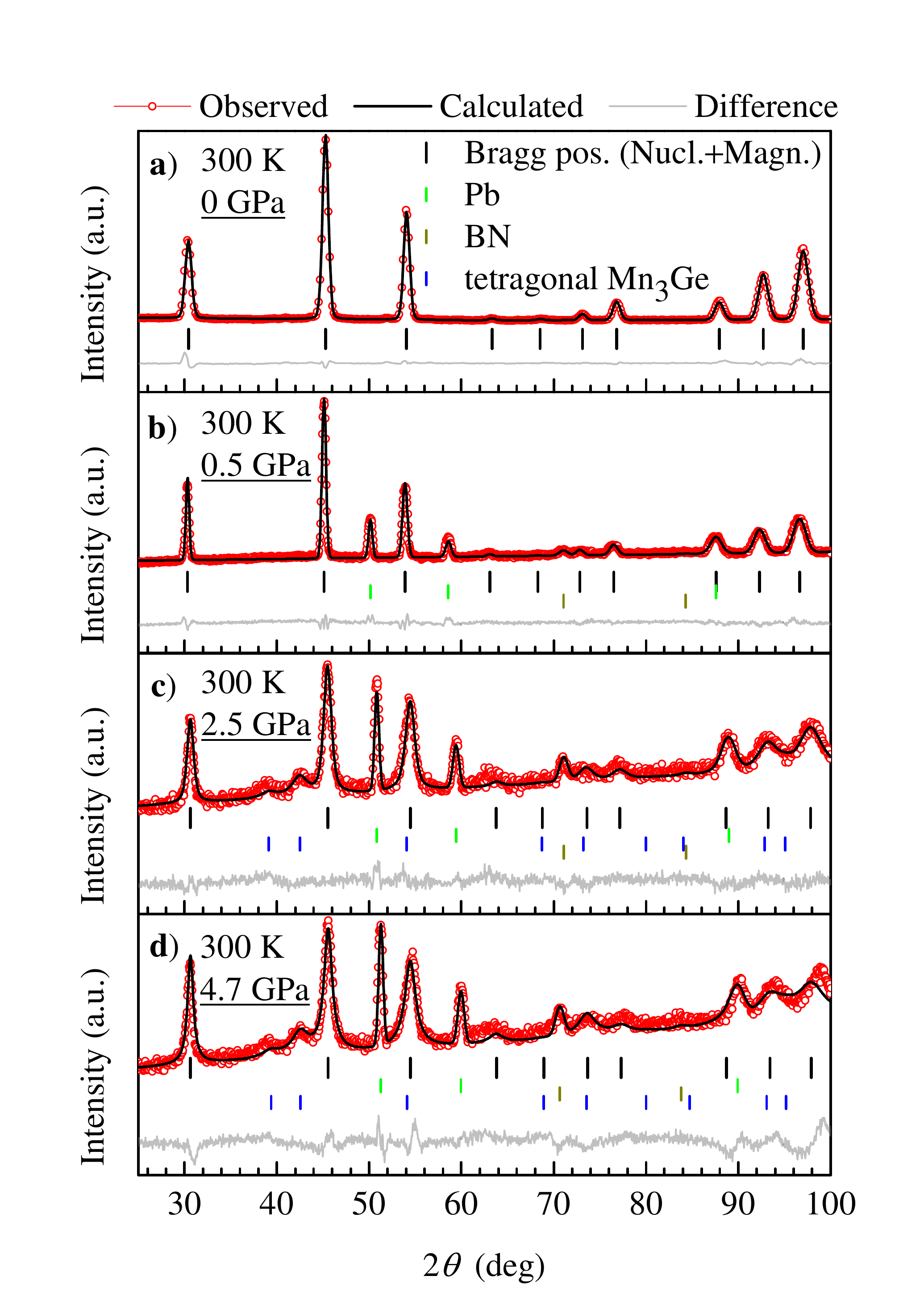}}
        \end{minipage}
        \caption{(color online). Set of diffraction patterns collected at different hydrostatic pressure at 300 K: (a) ambient data taken at the E6 diffractometer; (b)--(d) results of measurements on the D20 diffractometer. The red circles are experimental data. Solid black line and solid gray line correspondingly denote the result of Rietveld refinement and the residue. Vertical ticks mark the position of the Bragg peaks for the main phase as well for the additional phases (see text for details).}
        \label{ris:fig2}
\end{figure}

As reported elsewhere, the stoichiometric composition Mn$_3$Ge does not form a stable phase. Therefore, a polycrystalline sample of Mn$_{3.2}$Ge, referred hereafter as Mn$_3$Ge, was prepared by induction melting of the corresponding ratio of pure Mn (ChemPUR, 99.99\%) and Ge (ChemPUR, 99.9999\%) elements. The ingot was thoroughly ground and consequently annealed at $850\,^{\circ}\mathrm{C}$. X-ray diffraction and energy-dispersive x-ray analysis confirmed that the sample is a single-phase compound of Mn$_3$Ge with less than 1\% impurities.

Powder neutron diffraction measurements in a wide-temperature range at ambient pressure were conducted at the E6 diffractometer at the HZB (Berlin, Germany) \cite{E6_1,E6_2,E6_3}. The sample was encapsulated in a vanadium cylinder and inserted into a cryofurnace. A monochromatic neutron beam with a wavelength of 2.447\phantom{ }\AA  \phantom{ }was used for measurements at (80--500)\phantom{ }K. Measurements under hydrostatic pressure were performed at the D20 diffractometer at the ILL (Grenoble, France) \cite{D20_1}. A standard Paris-Edinburgh pressure cell with a cryostat was used to control the applied pressure. A small amount of Pb powder, which served as the standard for determining the on-sample pressure \cite{Pb}, was added to the sample placed in a Zr-Ti (null-scattering alloy) gasket. An ethanol-methanol mixture was used as a pressure-transmitting medium. The neutron beam was monochromized to 2.41\phantom{ }\AA  \phantom{ }by PG $(002)$ reflection. The powder patterns were collected on stepwise increase of pressure in the range of (0.5--5)\phantom{ }GPa at 300\phantom{ }K. At a few selected values of pressure the temperature scans from 80\phantom{ }K to 300\phantom{ }K were recorded. First, the sample was cooled to low temperature, then the applied hydrostatic pressure was maintained. The process of heating the sample led to a temperature-dependent offset in pressure (up to approximately 10\%) that was determined and taken into account. All the collected data have been analysed by Rietveld refinement method using the FullProf software \cite{FullProf}.

A typical powder neutron diffraction pattern is shown in Fig. \ref{ris:fig1}. The intensity of the peaks in the low-temperature pattern differs significantly from the pattern recorded in the paramagnetic phase by the growing intensity of the first 3 strongest reflections: $(100), (101)$ and $(110)$. The magnetic Bragg peaks appear on top of the nuclear reflections, as expected for a $\textbf{k} = 0$ magnetic structure. The $(200)$ peak, which is a weak nuclear reflection with $|F(200)|/|F(110)| = 0.116$, remains unchanged in the AFM phase as it is forbidden for the plain triangular structure.

Diffraction patterns obtained under hydrostatic pressure at D20 are shown in Fig. \ref{ris:fig2}(b--d) and compared to the data (a) taken at ambient pressure at E6. As can be seen, the effect of pressure results in a certain redistribution of intensity between the strong and the weak Bragg peaks. There are 3 additional phases with the Bragg peaks marked by the corresponding vertical lines along with the main phase of the Mn$_3$Ge compound. The second line (green ticks) shows Pb with two strong reflections at $2\theta \approx 50^{\circ}$ and $60^{\circ}$. The positions of the Bragg peaks of Pb were used to refine the lattice constant of the element, which allows us to calculate on-sample pressure for the given conditions if the equation of state for Pb is known \cite{Pb}. The dark yellow tick in the third line marks a spurious Bragg peak at $2\theta \approx 71^{\circ}$ coming from the boron nitride anvils of the Paris-Edinburgh pressure cell due to the incomplete absorption. The peak has a constant intensity throughout the whole set of data, however it is more pronounced at high pressures because the overall intensity from the sample in the gasket is reduced by the closing gap between the squeezed anvils, thus making it more visible on the relative scale. The last phase (blue ticks) is represented by the tetragonal polymorph of Mn$_3$Ge, which was absent before pressure was applied. Hexagonal Mn$_3$Ge is a metastable phase at $T < 953$ K, however the transition to the stable tetragonal phase does not occur unless the sample is annealed at sufficiently high temperature for a long time. The high barrier of hexagonal to tetragonal polymorph transformation seems to be noticeably lowered with high pressure. The tetragonal phase becomes visible as an impurity at $P > 1$\phantom{ }GPa and reaches as much as 15\% of the weight of the whole sample at $P = 4$\phantom{ }GPa. All the present phases were taken into account in the pattern refinement.

\section{Spin canting under hydrostatic pressure}\label{sec:III}

\begin{figure}[t]
        \begin{minipage}{0.99\linewidth}
        \center{\includegraphics[width=1\linewidth]{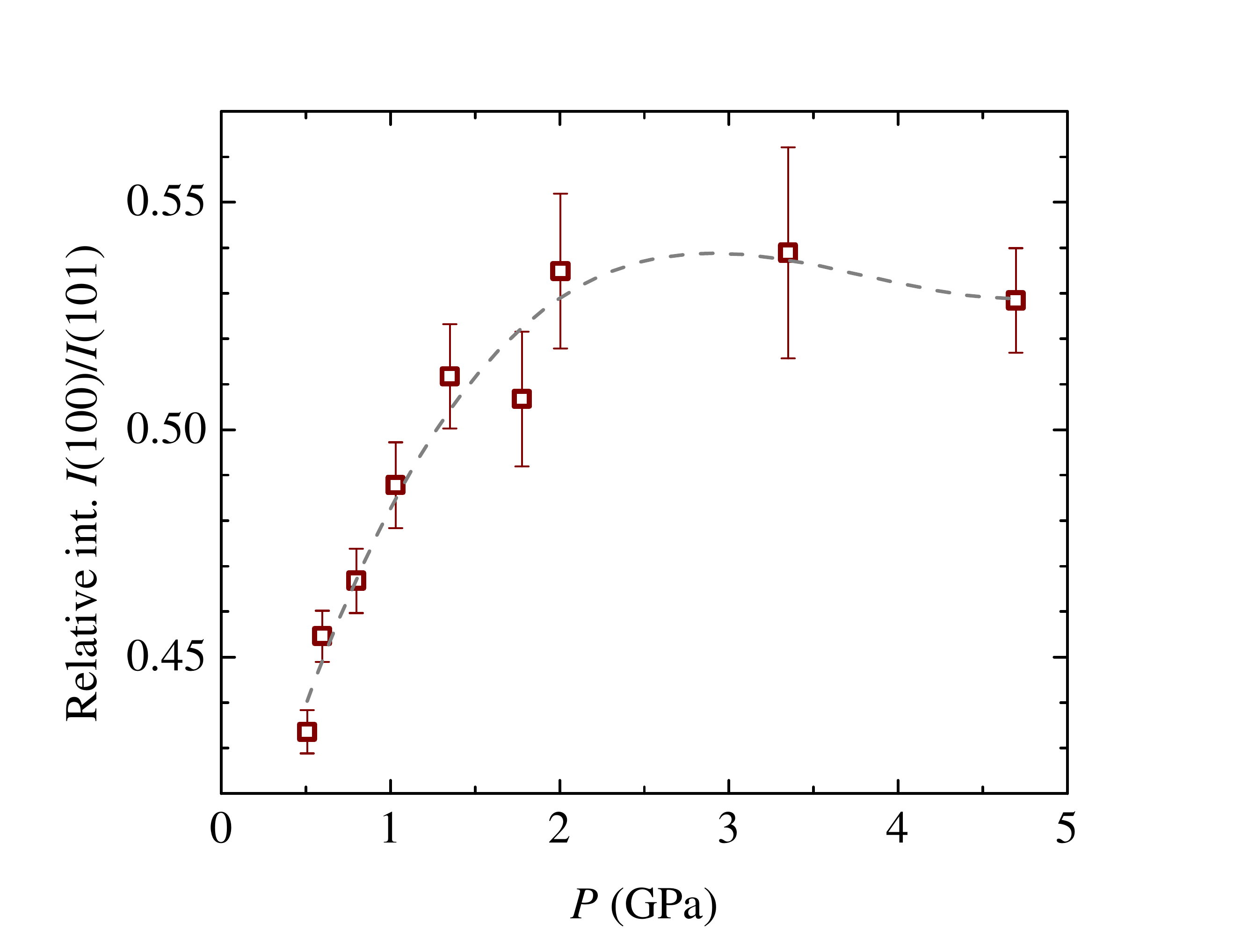}}
        \end{minipage}
        \caption{(color online). The change in the relative intensity of two strong nuclear and magnetic Bragg peaks (100) and (101) at different pressure. Dashed line is guide for the eyes.}
        \label{ris:fig1s}
\end{figure}

\begin{figure}[t]
        \begin{minipage}{0.99\linewidth}
        \center{\includegraphics[width=1\linewidth]{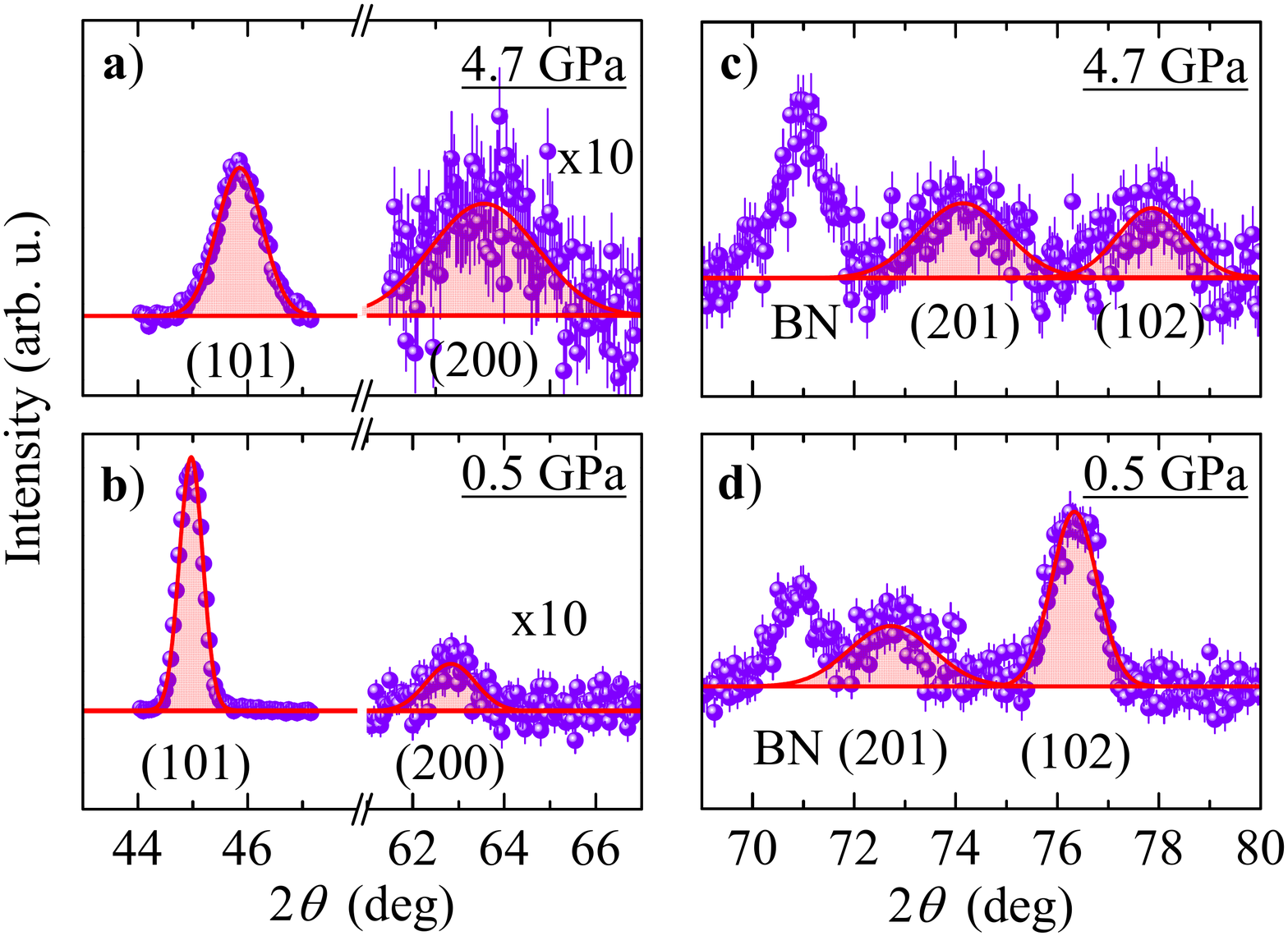}}
        \end{minipage}
        \caption{(color online). (a)--(b) The intensity of the $(200)$ Bragg peak compared with the (101) peak for high (a) and low (b) pressure. (c)--(d) the relative change between (201) and (102) reflections for high (c) and low (d) pressure. Solid lines are a Gaussian fit.}
        \label{ris:fig2p5}
\end{figure}

\begin{figure}[h]
        \begin{minipage}{0.99\linewidth}
        \center{\includegraphics[width=1\linewidth]{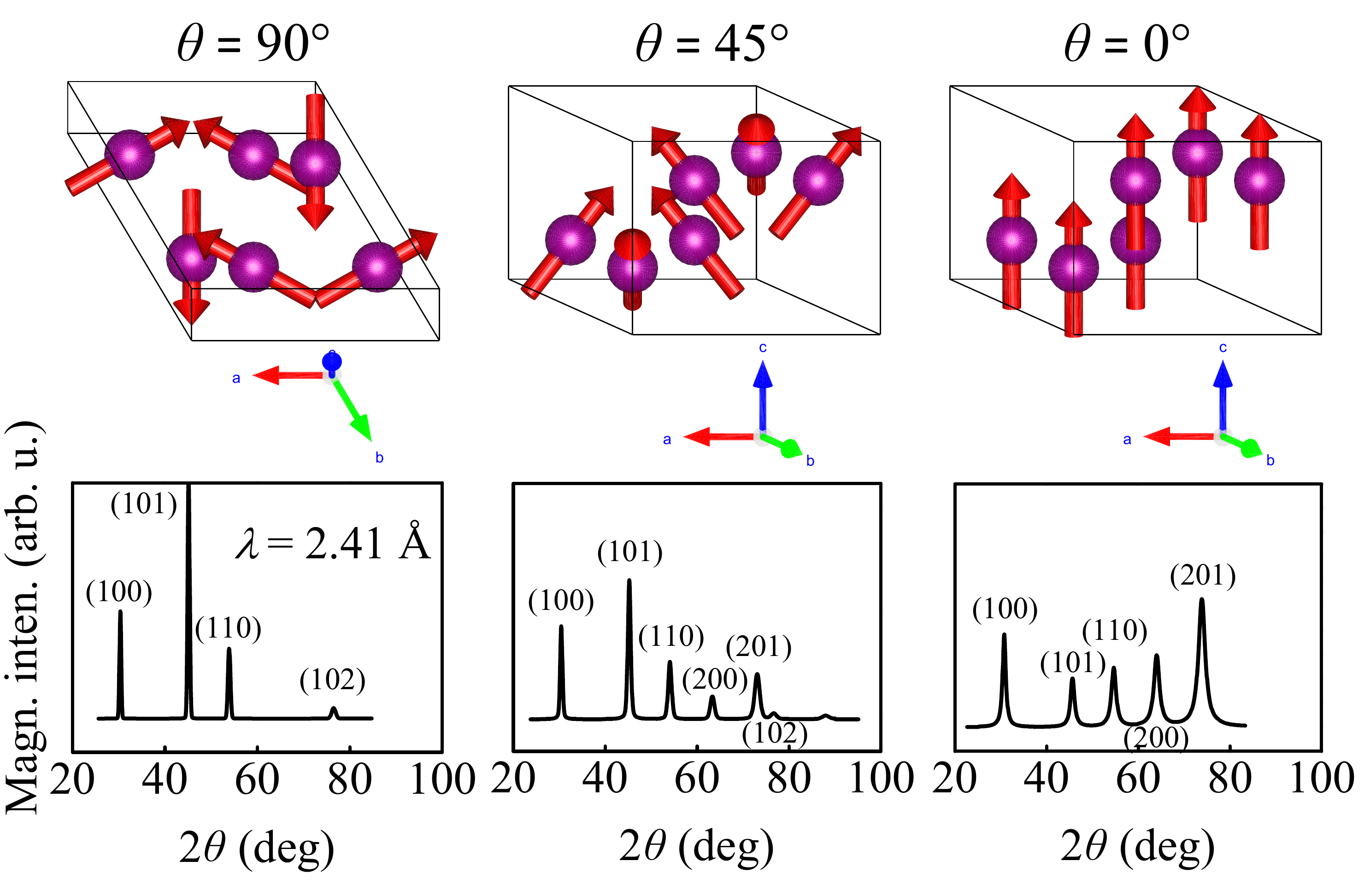}}
        \end{minipage}
        \caption{(color online). The inverse triangular 120$^{\circ}$-magnetic structure of Mn$_3$Ge shown together with the simulated magnetic contribution to the neutron diffraction pattern for different canting angles. The leftmost figure is a plain structure found to be the ground state at ambient pressure. The magnetic model discussed in the text is shown in the middle. The $\theta$ angle, which is counted from the $c$-axis,  denotes an uniform out-of-plane spin canting. The rightmost structure is a fully-polarized FM order.}
        \label{ris:fig0}
\end{figure}

\begin{figure}[t]
        \begin{minipage}{0.99\linewidth}
        \center{\includegraphics[width=1\linewidth]{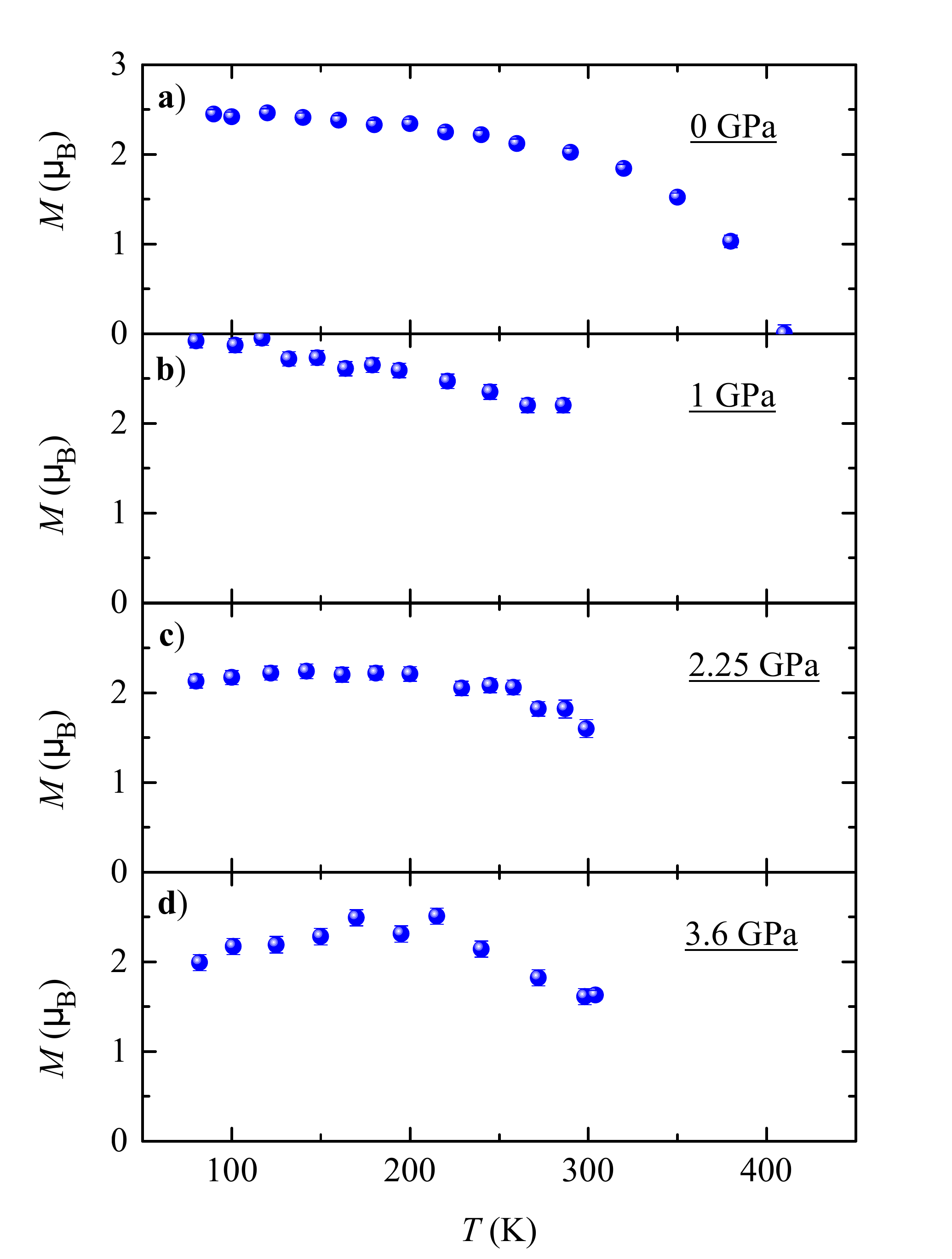}}
        \end{minipage}
        \caption{(color online). The refined value of the magnetic moment per Mn atom as a function of temperature for different pressure.}
        \label{ris:fig3}
\end{figure}

\begin{figure}[t]
        \begin{minipage}{0.99\linewidth}
        \center{\includegraphics[width=1\linewidth]{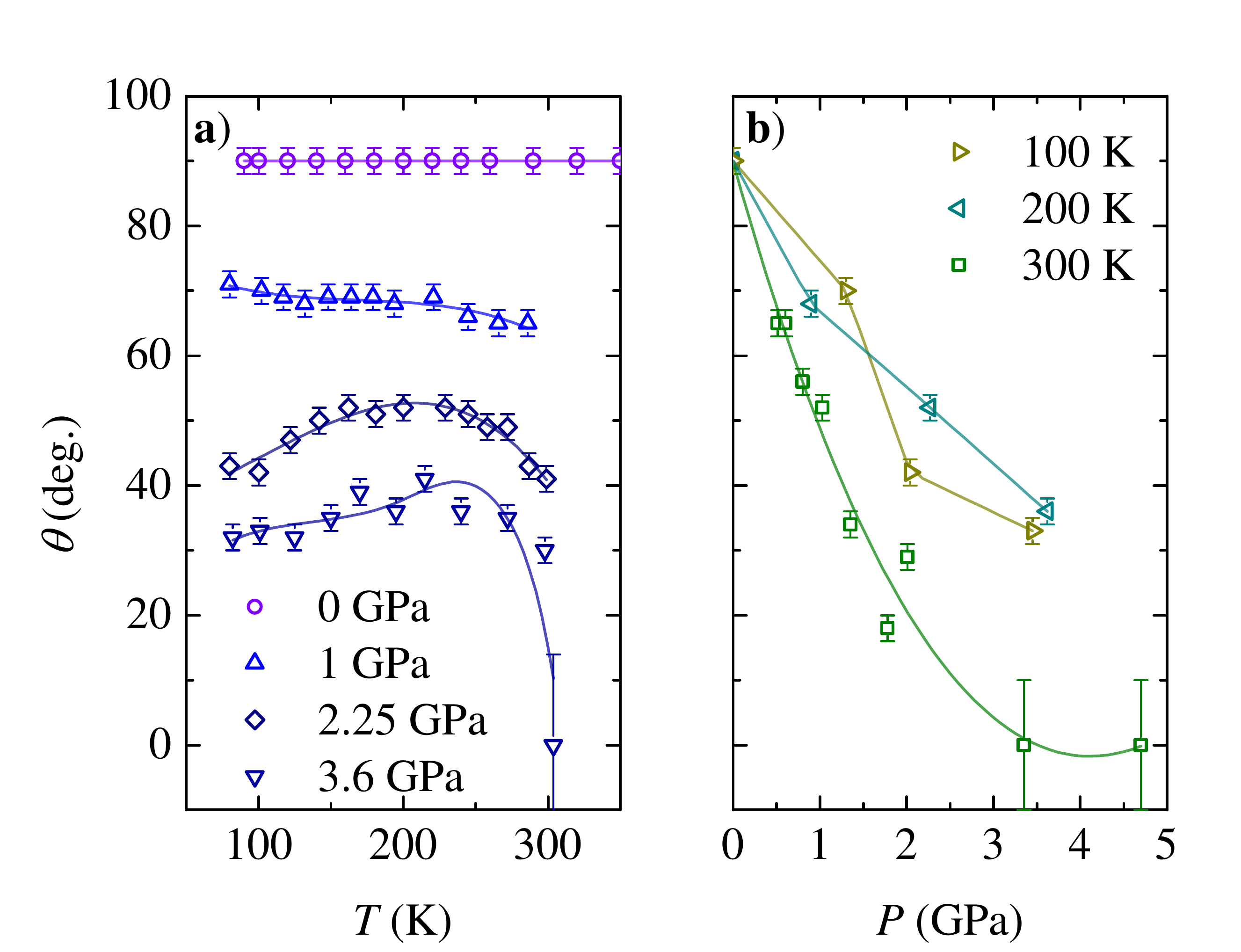}}
        \end{minipage}
        \caption{(color online). The refined value of the out-of-plane spin canting angle as a function of temperature for different pressure (a) and a function of pressure for different temperatures (b). Solid lines are guides for the eyes.}
        \label{ris:fig4}
\end{figure}

The intensities of different magnetic peaks were analysed in details. The (101) Bragg peak, which is the strongest peak by both nuclear and magnetic contributions, decreases in the intensity with applied pressure as seen on the background of the two other strong reflections (100) and (110). The indicative comparison of the (100) and (101) is demonstrated in Fig. \ref{ris:fig1s} as the relative intensity $I(100)/I(101)$ versus pressure at room temperature. The quantity $I(100)/I(101)$ yields a value of $\approx$ 0.5 in the paramagnetic state, thus giving a purely nuclear structure factor. Because the (101) is also stronger than the (100) in the magnetically ordered phase, the relative intensity is further imbalanced to smaller value. Upon applied pressure, the ratio $I(100)/I(101)$ monotonically increases in the pressure range of up to 2\phantom{ }GPa and reaches the magnitude of $\approx$ 0.53, whereas it stays almost the same between 2\phantom{ }GPa and 4.7\phantom{ }GPa. The change exceed the paramagnetic ratio and cannot be accounted for by a simple suppression of the magnetic order (gradual vanishing of the ordered magnetic moment). Such a behaviour of the strongest magnetic peaks suggests a change in the magnetic structure, for example, a spin canting. However, the analysis of the other reflections present in the pattern, including weak Bragg peaks, is essential for identification of the correct model of the canted magnetic structure.

Figures \ref{ris:fig2p5}(a)--(b) demonstrate the relative change between the strongest magnetic and nuclear (101) reflection and the weak nuclear (200) Bragg peak, which is magnetically forbidden for the planar AFM structure but allowed in the case of a collinear FM component along the $c$-axis. The profiles of the peaks were fitted with a Gaussian function that yields the change of the ratio of integral intensity (the area of the peak) between (200)/(101) from 0.05 at 0.5\phantom{ }GPa to 0.21 at 4.7\phantom{ }GPa. The relative change in another pair of the Bragg peaks is shown in Fig.  \ref{ris:fig2p5}(c)--(d). The (201) reflection is also forbidden for the coplanar structure and contains pure nuclear intensity at the ambient pressure. Similarly to the (200) Bragg peak, the (201) does not vanish the structure factor of the FM order. On the contrary, the closely placed (102) reciprocal space point has no magnetic intensity if the spins are fully aligned along the $c$-axis. As was found from the separate Gaussian fit of (201) and (102) for low and high pressures, the relative intensity $I(201)/I(102)$ changes from 0.55 to to 1.38 between 0.5 and 4.7\phantom{ }GPa. The observed redistribution of the intensity indicates the change in the magnetic ordering.

The model of magnetic phase with magnetic moments in spherical mode \cite{FullProf} was applied to study the evolution of the magnetic structure of Mn$_3$Ge under pressure. The model assumes 6 values of the azimuthal angle $\phi$ for each of the Mn atoms in the unit cell that form an inverse triangular 120$^{\circ}$ spin structure. The free parameters are the magnitude of magnetic moment per Mn atom and the polar angle $\theta$ that describes the out-of-plane uniform canting of magnetic moments. In this case the scale factor for the magnetic phase was kept equal to the scale factor of the corresponding crystallographic phase. The best fit for $\theta$ for ambient pressure was found to be $(90 \pm 5)^{\circ}$, unchanged with temperature, in agreement with previous reports \cite{N1,N2,N3,N4,N5}. The change in the magnetic structure is schematically drawn in Fig. \ref{ris:fig0} along with the corresponding simulated magnetic contribution to the neutron powder diffraction pattern. As can be seen, the spin canting results in the redistribution of the intensity between the strong (100), (101) and (110) reflections to the (200) and (201) reflections, which are magnetically forbidden for the coplanar AFM structure.

The refined size of the magnetic moment per Mn atom at ambient pressure is shown in Fig. \ref{ris:fig3} (a). The value shows no anomalies, monotonically decreases with temperature and eventually vanishes between 377 and 407 K, close to the reported $T_{\text N}$ \cite{E1,E2,E3}. The extrapolated to zero temperature magnetic moment is 2.52 $\mu_{\text B}$, the same value was obtained in \textit{ab initio} calculations in Ref. \cite{T21}. The effect of pressure on the magnetic moment is shown in Figs. 6(b)--6(d). It is slightly enhanced at a moderate pressure of 1\phantom{ }GPa to the extrapolated value of 2.86 $\mu_{\text B}$. Seemingly, the $T_{\text N}$ remains well above 300\phantom{ }K for pressures up to 5\phantom{ }GPa. The magnetic moment decreases back to $\approx$ 2.34 $\mu_{\text B}$ at 2.25\phantom{ }GPa and keeps the same unchanged value at 3.6\phantom{ }GPa with possible small maximum at 200\phantom{ }K. One may conclude that the electronic band splitting is not affected by pressure in this range. Neither does the pressure change of lattice constants affect the magnetic exchange energies, since the ordering temperature does not show any noticeable change.

The angle of out-of-plane canting $\theta$ evolves in the $(T, P)$-parameter space  as shown in Fig. \ref{ris:fig4}. As was previously mentioned, there is no canting at ambient pressure ($\theta = 90^{\circ}$). The spins uniformly move out of plane to $\theta \simeq 70^{\circ}$ as shown for $P = 1$\phantom{ }GPa, then to $\theta \simeq 50^{\circ}$ and $30^{\circ}$ for $P = $2.25 and 3.6\phantom{ }GPa, respectively. In general, the canting angle $\theta$ depends mainly on pressure rather than on temperature but tends to reduce faster at 300\phantom{ }K than at lower temperatures as illustrated in Fig. \ref{ris:fig4} (b). The magnetic structure becomes fully polarized with $\theta = 0^{\circ}$ at approximately 3.4\phantom{ }GPa at 300\phantom{ }K. The system remains ferromagnetic at 300\phantom{ }K at 4.7\phantom{ }GPa with magnetic moment of $(1.8 \pm 0.1) \mu_{\text B}$.

\section{Magnetoelastic phenomena}\label{sec:IV}

\begin{figure*}
\includegraphics[width=\linewidth]{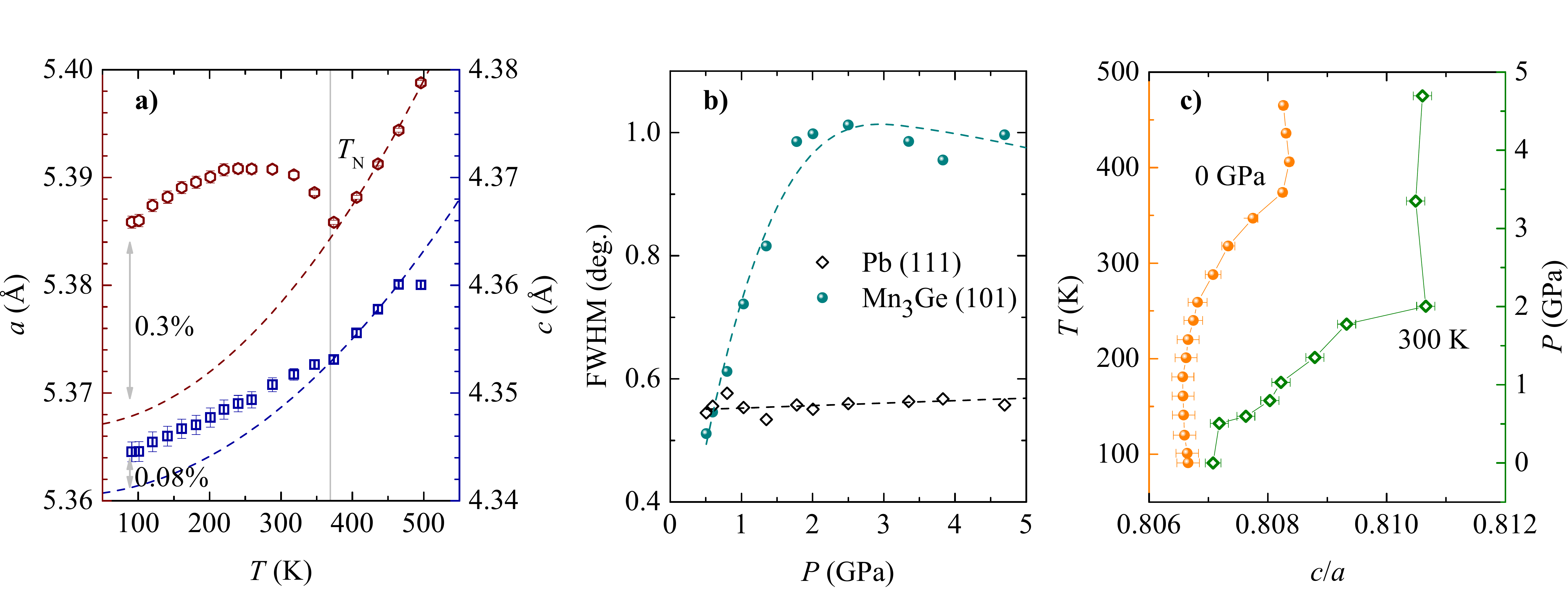}\vspace{3pt}
        \caption{(color online). The magnetoelastic effect: (a) Temperature change of the lattice parameters $a$ (red hexagons, left scale) and $c$ (blue squares, right scale). An anomalous thermal expansion can be seen in the AFM phase. Dashed lines are simple approximation for paramagnetic thermal expansion (see text for details); (b) Change of the width (FWHM) of Bragg peaks of Mn$_3$Ge as compared to the Pb reference for different pressure. Dashed lines are guides for the eyes. (c) The $c/a$ ratio as a function of temperature at ambient pressure (left scale) and pressure at 300\phantom{ }K (right scale).}
        \label{ris:fig5}
\end{figure*}

We now discuss the possible explanation for the gradual out-of-plane spin canting under hydrostatic pressure. The ground state of Mn$_3$Ge, i.e. a plain triangular structure, can be well understood as the result of the nearest neighbour AFM exchange interaction. The 120$^{\circ}$ spin order is therefore a result of geometrical frustration of the kagom\'e lattice. The magnetic propagation vector remains degenerate between $\textbf{k} = 0$ and $\textbf{k} = (1/3,1/3,0)$ (so-called $\sqrt{3}\times\sqrt{3}$ order). Depending on the sign and absolute value, the next-nearest neighbour in-plane interaction lifts this degeneracy \cite{kagom1,kagom2,kagom3,kagom4,kagom5,kagom6}. This 2D interaction scheme can be extended to the 3D structure of closely-packed kagom\'e layers, as it is the case for Mn$_3$Ge. If taken into account, the sign of exchange between the triangles of atoms on adjacent layers will determine whether the related by inversion spins prefer parallel or antiparallel alignment.

However, there is no obvious exchange interaction scheme that would lead to the uniform out-of-plane spin canting with a balance angle $\theta$. If the dominant nearest-neighbour interaction changed its sign, that would result in a phase transition from the 120$^{\circ}$ order to a collinear FM state with no intermediate canting. Either single-ion anisotropy or the Dzyaloshinskii-Moriya interaction could induce the spin canting, but they are expected to be an order of magnitude weaker than the exchange interaction in $3d$ metals and, thus, unlikely to be considered as the mechanism of spin rotation at large angles. An alternative scenario can include effects of magnetoelastic phenomena found to be strong for many magnets, regardless of the particular type of their magnetic ordering (\cite{Rev21,Rev22} and refs. therein).

Fig. \ref{ris:fig5}(a) shows the temperature change of the refined lattice parameters $a$ and $c$ in the range of (80--500)\phantom{ }K at ambient pressure. Both lattice parameters obey a conventional thermal expansion above $T_{\text N}$ as can be approximated by a simple function $a(T) = 5.367+1.3 \cdot 10^{-7}T^2$ and $c(T) = 4.341+0.9 \cdot 10^{-7}T^2$ respectively for $a$ and $c$. As temperature passes below $T_{\text N}$ on cooling, a large negative thermal expansion can be observed in the $a$ lattice constant and a small yet visible change of the slope in the $c$ parameter. The thermal expansion in the absence of magnetism is not known but can be very roughly extrapolated from the paramagnetic region \cite{Tapan1,Tapan2,Tapan3,Tapan4}. This yields an estimation for the strain $\Delta a/a = 3\cdot 10^{-3}$ and $\Delta b/b = 0.8\cdot 10^{-3}$ accumulated due to spontaneous magnetostriction. The dominant in-plane strain implies that it is directly related to the in-plane non-collinear AFM structure. The plain non-collinear spin texture therefore causes an effective negative pressure. The resulting volumetric spontaneous magnetostrain at 90\phantom{ }K is estimated to be $\simeq$ 0.6\%. For comparison, a spontaneous magnetostrain of 3.1\% was reported for the hard magnet Tb$_2$Fe$_{14}$B as the largest ever observed \cite{ME1,ME2}. As an example of a non-collinear AFM, $\alpha$-Mn shows the same effect of only 0.13\% \cite{ME3}.

Once the external hydrostatic pressure is applied, the system experiences competing influences that induce a non-uniform strain on the crystal lattice, seen as a broadening of the Bragg peaks of Mn$_3$Ge. The FWHM is depicted in Fig. \ref{ris:fig5}(b) as a function of pressure and compared with the width of peaks of the Pb reference. As can be seen, the peaks of Pb remain unchanged and indicate that the applied pressure is to a great extent hydrostatic (the width is limited by the resolution of the diffractometer). In turn, the peaks of Mn$_3$Ge become twice broader in the region from 0.5\phantom{ }GPa to 2\phantom{ }GPa and seem to preserve the width after 2\phantom{ }GPa. We note, that a similar effect could in principle be seen if Mn$_3$Ge had much greater compression (smaller bulk modulus) than Pb. If this is assumed, pressure on the sample that varies within $\delta P$ would give larger spread of the lattice constant for Mn$_3$Ge than for Pb and would be specifically seen on the former. However, the bulk modulus of Mn$_3$Ge is $\sim$ 1.5 times greater than the bulk modulus of Pb ($\Delta V/V \simeq 0.09$) for the pressure range (0--8.7)\phantom{ }GPa at 300\phantom{ }K. Thus, the drastic change in the FWHM of Mn$_3$Ge must be attributed to a physical change in the system.

The magnetoelastic effect can also be illustrated by a change in the $c/a$ ratio plotted in Fig. \ref{ris:fig5}(c) as a function of temperature for ambient pressure and as a function of pressure for 300\phantom{ }K. The $c/a$ ratio is constant in the paramagnetic phase and starts decreasing below the AFM transition until $\sim$ 250\phantom{ }K where it stops changing for the lower temperature region. On the contrary, pressure leads to an increase in the $c/a$ back to the paramagnetic value at $\sim$ 1\phantom{ }GPa, and it continues to increase between 1 and 2\phantom{ }GPa. The total pressure-induced uniaxial elongation is roughly equal to the contraction caused by the plain triangular spin configuration at ambient pressure. The elongation in the $c$ axis is likely to mean that spontaneous magnetostriction acquires a reverse effect to the unit cell. In other words, the magnetostrain in the AFM phase under pressure is greater along the $c$-axis. As one can see, the latter correlates with the out-of-plane rotation of the magnetic moments.

\section{Conclusions}\label{sec:V}

To summarize, we have conducted powder neutron diffraction experiments on the non-collinear AFM Mn$_3$Ge under hydrostatic pressure. The application of pressure  up to 5\phantom{ }GPa causes a gradual change from the non-collinear triangular  magnetic structure to a uniformly canted non-collinear triangular structure and a successive change to the collinear FM structure. Diffraction measurements in a wide temperature range at ambient pressure revealed a sizeable spontaneous magnetostriction below $T_{\text{N}}$. The magnetostrain is mainly accumulated in the basal plane of the hexagonal structure in accordance with the coplanar triangular magnetic structure. Pressure leads to a change in the $c/a$-ratio, possibly indicating a crossover in the distortion of the unit cell from the in-plane to the out-of-plane elongation. These distortions, or the magnetostrain, reasonably correlate with the out-of-plane canting of the magnetic structure.

The observed change in the non-collinear magnetic structure under pressure might cause changes in the Berry phase and the Berry curvature and, consequently, lead to to change in the associated transport properties such as the AHE or anomalous Nernst effect \cite{Nerst}. The fact that the spin structure can be changed in a gradual fashion opens an opportunity  to ``tune" the Berry phase in a desired way, which is important for the study of different magnetotransport phenomena or for spintronic applications.

We argue that similar changes in the magnetic structure under pressure may be expected for the related compounds Mn$_3X$, where $X$ = Sn, Ga, Ir, Rh or Pt.

\section*{Acknowledgments}\label{sec:V}
We thank S. E. Nikitin for stimulating discussions. A.S.S. acknowledges support from the International Max Planck Research School for Chemistry and Physics of Quantum Materials (IMPRS-CPQM). S.S. thanks Science and Engineering Research Board of India for the Ramanujan Fellowship. The work at the TU Dresden was funded by the German Research Foundation (DFG) in the framework of the Collaborative Research Center SFB 1143 (project C03) and the Priority Program SPP 2137 ``Skyrmionics''.


%
\end{document}